\documentclass[aps,prl,
twocolumn,nofootinbib,
superscriptaddress,preprintnumbers]{revtex4-1}

\usepackage{amsmath,amsfonts,amssymb}
\usepackage[dvipsnames,x11names]{xcolor}
\usepackage{graphicx,bm,color,subfigure,hyperref,xspace,multirow}
\hypersetup{ 
    pdfnewwindow=true,   
    colorlinks=true,     
    linkcolor=blue,      
    citecolor=blue,      
    filecolor=blue,      
    urlcolor=blue        
}

\usepackage{soul}

\newcommand{\jlab}{Thomas Jefferson National Accelerator Facility, Newport News, VA 23606, USA}
\newcommand{\ceem}{Center for  Exploration  of  Energy  and  Matter, Indiana University, Bloomington,  IN 47403, USA}
\newcommand{\indiana}{Physics Department, Indiana University, Bloomington, IN 47405, USA}
\newcommand{\ectstar}{European Centre for Theoretical Studies in Nuclear Physics and Related
Areas (ECT$^*$) and Fondazione Bruno Kessler,
I-38123 Villazzano (TN), Italy}
\newcommand{\icn}{Instituto de Ciencias Nucleares, 
Universidad Nacional Aut\'onoma de M\'exico, Ciudad de M\'exico 04510, Mexico}
\newcommand{\genova}{INFN Sezione di Genova, Genova, I-16146, Italy}
\newcommand{\cern}{CERN, 1211 Geneva 23, Switzerland}
\newcommand{\ucm}{Departamento de F\'isica Te\'orica, Universidad Complutense de Madrid, 28040 Madrid, Spain}

\usepackage{xspace}
\usepackage{graphicx}
\usepackage{dcolumn}
\usepackage{bm}
\usepackage{hyperref}
\usepackage{braket}
\usepackage{units}

\usepackage{dsfont}
\usepackage{todonotes}
\usepackage{multirow}

\newcommand{\lhcb}{LHCb\xspace}

\newcommand{\ie}{{\it i.e.}\xspace}
\newcommand{\abs}[1]{\ensuremath{|#1|}}
\DeclareMathOperator{\im}{Im}
\DeclareMathOperator{\re}{Re}
\newcommand{\SigmaD}{\ensuremath{\Sigma_c^+ \bar{D}^0}\xspace}
\newcommand{\Pc}{\ensuremath{P_c(4312)^+}\xspace}

\newcommand{\jpsip}{\ensuremath{J/\psi\,p}\xspace}
\usepackage{array}   
\newcolumntype{L}{>{$}l<{$}} 
\newcolumntype{R}{>{$}r<{$}}
\newcolumntype{C}{>{$}c<{$}}
\usepackage{color}
\usepackage{xcolor}
\usepackage{slashed}

\newcommand{\mevnospace}{\ensuremath{{\mathrm{\,Me\kern -0.1em V}}}}
\newcommand{\gevnospace}{\ensuremath{{\mathrm{\,Ge\kern -0.1em V}}}}
\newcommand{\tevnospace}{\ensuremath{{\mathrm{\,Te\kern -0.1em V}}}}
\newcommand{\mev}{\mevnospace\xspace}
\newcommand{\gev}{\gevnospace\xspace}

\newcommand{\mytitle}[1]{\vspace{.5cm}{\em #1.---}}

\begin{document}
\preprint{JLAB-THY-19-2921}

\title{Interpretation of the LHCb \Pc Signal}

\author{C.~\surname{Fern\'andez-Ram\'irez}}
\email{cesar.fernandez@nucleares.unam.mx}
\affiliation{\icn}
\author{A.~\surname{Pilloni}}
\email{pillaus@jlab.org}
\affiliation{\ectstar}
\affiliation{\genova}
\author{M.~Albaladejo}
\affiliation{\jlab}
\author{A.~\surname{Jackura}}
\affiliation{\ceem}
\affiliation{\indiana}
\author{V.~\surname{Mathieu}}
\affiliation{\jlab}
\affiliation{\ucm}
\author{M.~Mikhasenko}
\affiliation{\cern}
\author{J.~A.~\surname{Silva-Castro}}
\affiliation{\icn}
\author{A.~P.~\surname{Szczepaniak}}
\affiliation{\jlab}
\affiliation{\ceem}
\affiliation{\indiana}

\collaboration{Joint Physics Analysis Center}
\noaffiliation


\begin{abstract} 
We study the nature of the new signal reported by LHCb in the \jpsip spectrum. Based on the $S$-matrix principles, we perform a minimum-bias 
analysis of the underlying reaction amplitude,  
focusing on the analytic properties that can be related to the microscopic origin of the \Pc peak.
By exploring several amplitude parametrizations, we find evidence for the attractive effect of the \SigmaD channel,
which is not strong enough, however, to form a bound state. 
\end{abstract}

\maketitle
\mytitle{Introduction}
   From first principles of QCD, it is still unknown why the vast majority of hadrons  appear to follow the valence quark model pattern proposed by Gell-Mann and Zweig~\cite{GellMann:1964nj}
   and Zweig~\cite{Zweig:1981pd}. The discovery of genuine multiquark states would be 
    a major milestone in the history of strong interactions. In recent years, several exotic candidates have been reported~\cite{Esposito:2016noz,Lebed:2016hpi,Guo:2017jvc,Olsen:2017bmm,Karliner:2017qhf,ali2019multiquark}. 
The observation by LHCb of a narrow peak at $4312\mev$ in the \jpsip  invariant mass distribution in the  $\Lambda_b^0 \to \jpsip\, K^-$ decay~\cite{Aaij:2019vzc} points to 
  yet another hidden charm  pentaquark. 
  A hint of this signal, labeled \Pc, was already visible in the earlier LHCb analyses, but it was statistically insignificant~\cite{Aaij:2015tga,Aaij:2016phn}. 
  The fact that such a narrow ($\sim 10\mev$) peak stands out in 
  what otherwise  appears to be a smooth background permits 
 a simple one-dimensional analysis, although determination of its quantum numbers will require the full six-dimensional 
amplitude analysis fitting both the energy and angular dependencies. 
    
The signal peaks approximately $5\mev$ below the \SigmaD threshold.
 It is often said that an enhancement in the proximity of a two-particle threshold is a manifestation of a hadron molecule composed of the two particles. A $J^P = 1/2^-$ \SigmaD molecule in the $4260$-$4300\mev$ region was indeed predicted in various 
 models~\cite{Wu:2010jy,Wu:2010vk,Wang:2011rga,Yang:2011wz,Xiao:2013yca,Yamaguchi:2017zmn}.
   However, this is not the only possibility. 
   Virtual states can be produced as well~\cite{Eden:1964zz}, for example by an attractive interaction that is not strong enough to bind a state,  as in neutron-neutron scattering~\cite{Hammer:2014rba}.
 Genuine compact pentaquark interpretations are also possible. A $3/2^-$ pentaquark was found in Ref.~\cite{Zhu:2015bba} at $4329\mev$. Compact diquark-diquark-antiquark states with  
       spin assignment $(1/2,3/2)^-$ at 
       $\sim4260\mev$, together with orbital excitations $(1/2,3/2)^+$  at $\sim4330\mev$,  were predicted in~\cite{Maiani:2015vwa}, and are compatible with a \Pc.\footnote{We recall that the compact pentaquark predictions rely on the preferred determination of the quantum numbers of the  $P_c(4380)^+$ and $P_c(4450)^+$, which might change when the two-state structure of the latter peak will be taken into account. The role of thresholds in multiquark states has been discussed in~\cite{Blitz:2015nra,Esposito:2016itg}. } 

These various interpretations of the \Pc signal are related to different analytic properties of the   $\Lambda_b^0 \to \jpsip \,K^-$ amplitude.
   In this Letter we investigate what can be concluded from the LHCb data on the \jpsip mass spectrum as far as the nature of the \Pc peak is concerned. 
   
\mytitle{Data and analysis of the \Pc region}
The mass and width of the $P_c(4312)^+$ as determined from the LHCb analysis have been compared with predictions and postdictions of several models~(see, for example, Refs.~\cite{Chen:2019asm,Chen:2019bip,Guo:2019fdo,Liu:2019tjn,Huang:2019jlf,Ali:2019npk,Giannuzzi:2019esi,Xiao:2019mvs,Shimizu:2019ptd,Guo:2019kdc,Xiao:2019aya}). In contrast, we follow here a minimally biased approach. We construct a reaction amplitude that respects the generic principles of the $S$-matrix theory, and fit directly the experimental $J/\psi\, p$ mass distribution. The $S$-matrix principles of unitarity and analyticity cannot fully determine the partial wave amplitudes, and  unless  the complete (infinite-dimensional, crossing symmetric) $S$-matrix is calculated, there will be 
 undetermined parameters. These encode specifics of the underlying QCD dynamics. We leave them to be determined by data, rather than by a given model. We fit the $\cos\theta_{P_c}$-weighted spectrum $dN/d\sqrt{s}$ 
     measured in Ref.~\cite{Aaij:2019vzc}, with $\sqrt{s}$   being the \jpsip    invariant mass, 
 and restrict the analysis to the $4250$-$4380\mev$ region where the \Pc is found. 
As a cross-check we also analyze the unweighted \jpsip spectrum in the same region, 
both with and without the $m_{Kp} > 1.9\gev$ cut.

\begin{figure*}
\centering
\includegraphics[width=.49\textwidth]{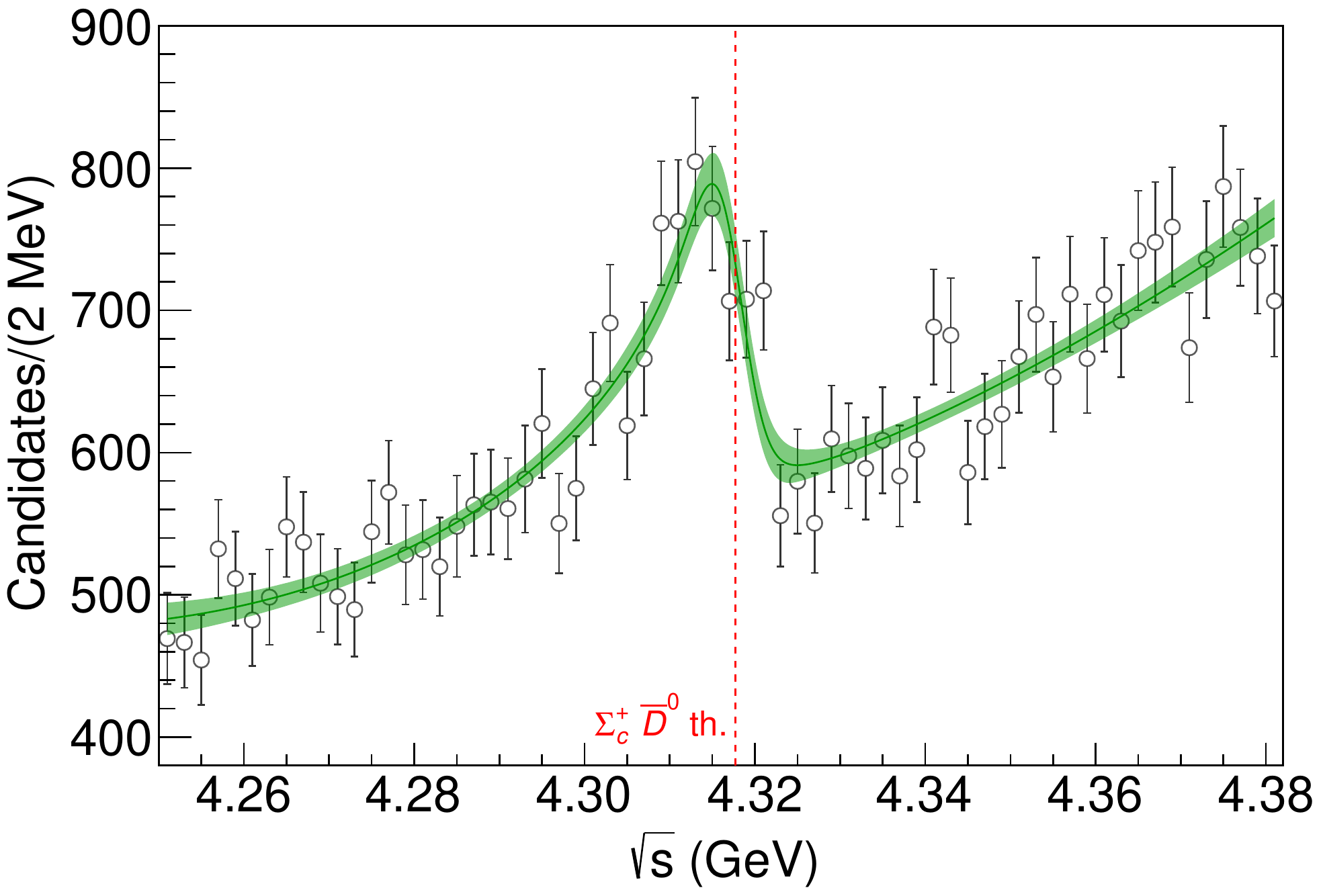}
\includegraphics[width=.49\textwidth]{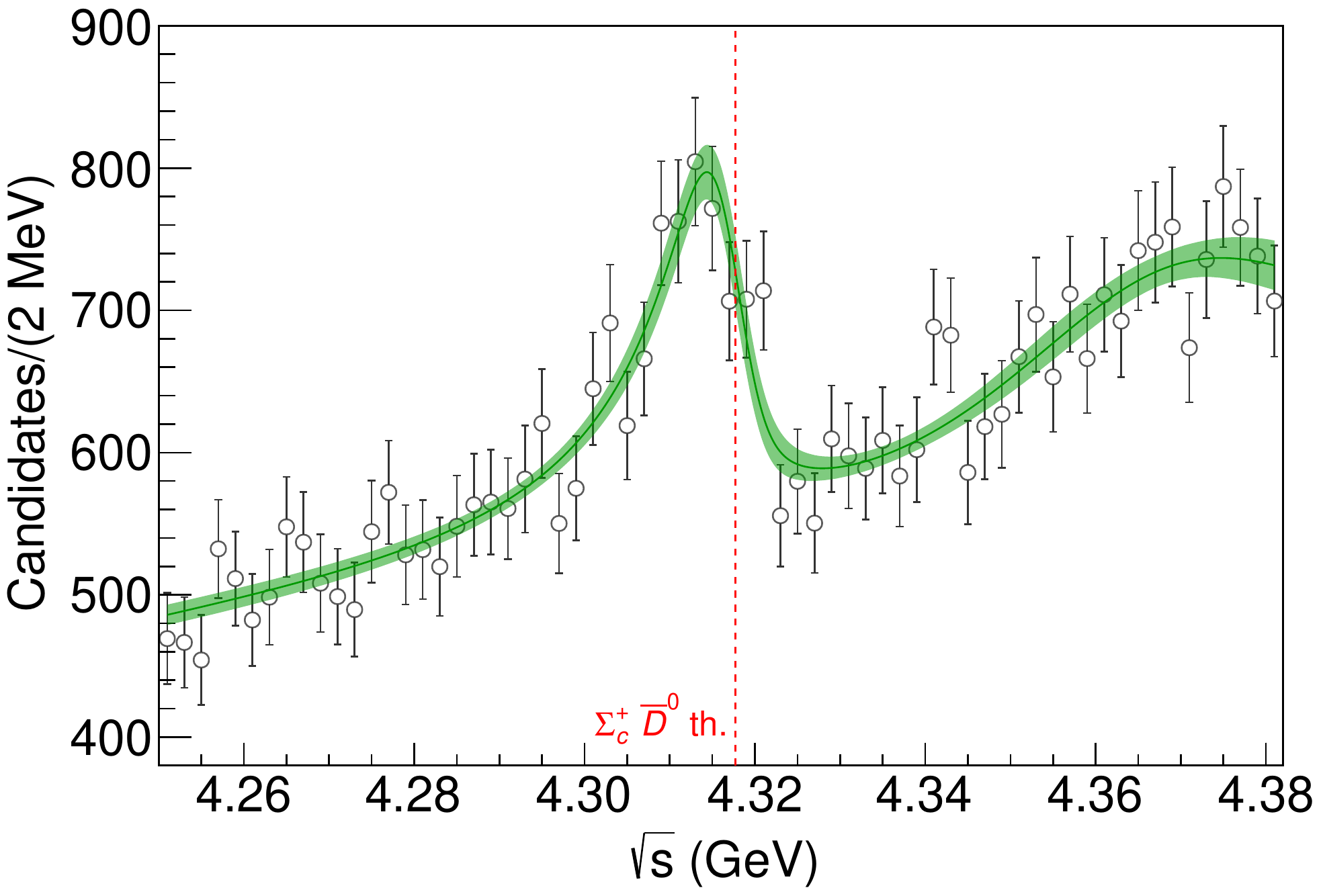}
\caption{
Fits to the $\cos\theta_{P_c}$-weighted \jpsip mass distribution from LHCb~\cite{Aaij:2019vzc} according to cases~A (left) and~B (right). 
The amplitude of case~A is expressed in the scattering length approximation, \ie $c_{ij}=0$ in Eq.~\eqref{eq:m}, and is able to describe either bound (molecular) or virtual states. The amplitude of case~B is given in the effective range approximation, \ie finite $c_{ii}$, and extends the description to genuine pentaquark states.
The solid line and green 
band show the result of the fit and the $1\sigma$ 
confidence level provided by the bootstrap analysis, respectively.
} \label{fig:fits}
\end{figure*}
As mentioned, the effect of the \SigmaD threshold looks prominent in data. We thus consider an amplitude which couples \jpsip (channel 1) and \SigmaD (channel 2). There is another nearby threshold, $6\mev$ above, which corresponds to the opening of the isospin partner,  $\Sigma^{++}_c D^-$ state. 
The \jpsip spectrum suggests this heavier threshold to be less 
important. We thus discuss the  two-channel case first, where the analytic properties are more transparent. We comment on the results of the three-channel fit farther below.  
The events distribution is given by 
\begin{equation}
\frac{dN}{d\sqrt{s}}= \rho(s) \left[ \abs{F(s)}^2   
 + B(s) 
\right],\label{1}
\end{equation} 
where $\rho(s)$ is the phase space factor. We assume that the \Pc signal has well-defined spin, \ie it appears in a single partial wave $F(s)$. The background $B(s)$ from all other partial waves is added incoherently, and parametrized with a linear polynomial.  
The amplitude $F(s)$ is a product of a 
  function $P_1(s)$ which provides
the production of $\jpsip\,K^-$,\footnote{The $P_1(s)$ function absorbs also the cross channel $\Lambda^*$ resonances projected into the same partial wave as \Pc. } and the $T_{11}(s)$ amplitude, which describes 
   the  $\jpsip\to\jpsip$ scattering,
\begin{equation} 
F(s) = P_1(s)\, T_{11} (s),\quad \left(T^{-1}\right)_{ij} = M_{ij} - i k_i \,\delta_{ij}, \label{eq:caseAB}
\end{equation} 
with $i,j =1,2$. Here  $k_i = \sqrt{s - s_i}$ with  $s_1 = (m_{\psi} + m_{p})^2$, $s_2 = (m_{\Sigma^+_c} + m_{\bar{D}^0})^2$ are the thresholds of the two channels. 
Although unitarity would prescribe to replace $k_i$ by the two-body phase space,
we approximated it by a 
 square root alone. This 
is fully consistent with the effective range expansion near threshold~\cite{Frazer:1964zz,Fernandez-Ramirez:2015tfa,Pilloni:2016obd}. We  stress that, since the \jpsip threshold is far away from the region of interest, this channel can effectively absorb all the other channels with distant thresholds.
In principle, one could also add the off-diagonal $P_2(s) T_{21}(s)$ term. This would not change the analytic properties, and would provide a nonzero value of $F(s)$ when $T_{11}(s)$ vanishes. The presence of a zero would be a relevant feature if no background were present, and in that case  $P_2(s) T_{21}(s)$ might be needed. 
In our case, we suppress such a term to reduce the number of free parameters.
For the real  symmetric  $2\times 2$ matrix $M(s)$ we use the first-order  effective range expansion 
\begin{equation} 
M_{ij}(s) = m_{ij}  - c_{ij} s, \label{eq:m}
\end{equation} 
 which is sufficient when considering the possibility of at most a single 
   threshold state (virtual or molecular)
   and a compact state~\cite{Frazer:1964zz}. In the single-channel case, this parametrization has often been discussed in the context of the Weinberg compositeness criterion~\cite{Weinberg:1965zz,Aceti:2012dd,Sekihara:2014kya,Guo:2015daa,Baru:2003qq,Guo:2019kdc}.
Any contribution from further singularities is smooth in the region of interest, and hence 
effectively incorporated in the background parameters.
The only exception is the triangle singularity,
already ruled out by LHCb as responsible for the signal~\cite{Aaij:2019vzc}.
  The function $P_1(s)$ is analytic in the data region, and, given the small mass range considered, it can 
  be parametrized with a first-order polynomial. For particle masses, we use the PDG values $m_{\Sigma_c^+}=2452.9\mev$ and $m_{\bar{D}^0}=1864.83\mev$~\cite{pdg}.
  Since the width of the $\Sigma_c^+$ is similar to the experimental resolution we neglect its effect.
  More details about the parametrizations and the fit results are in the Supplemental Material~\cite{pcjpaclink}.

\begin{figure*}
\centering
\includegraphics[width=.49\textwidth]{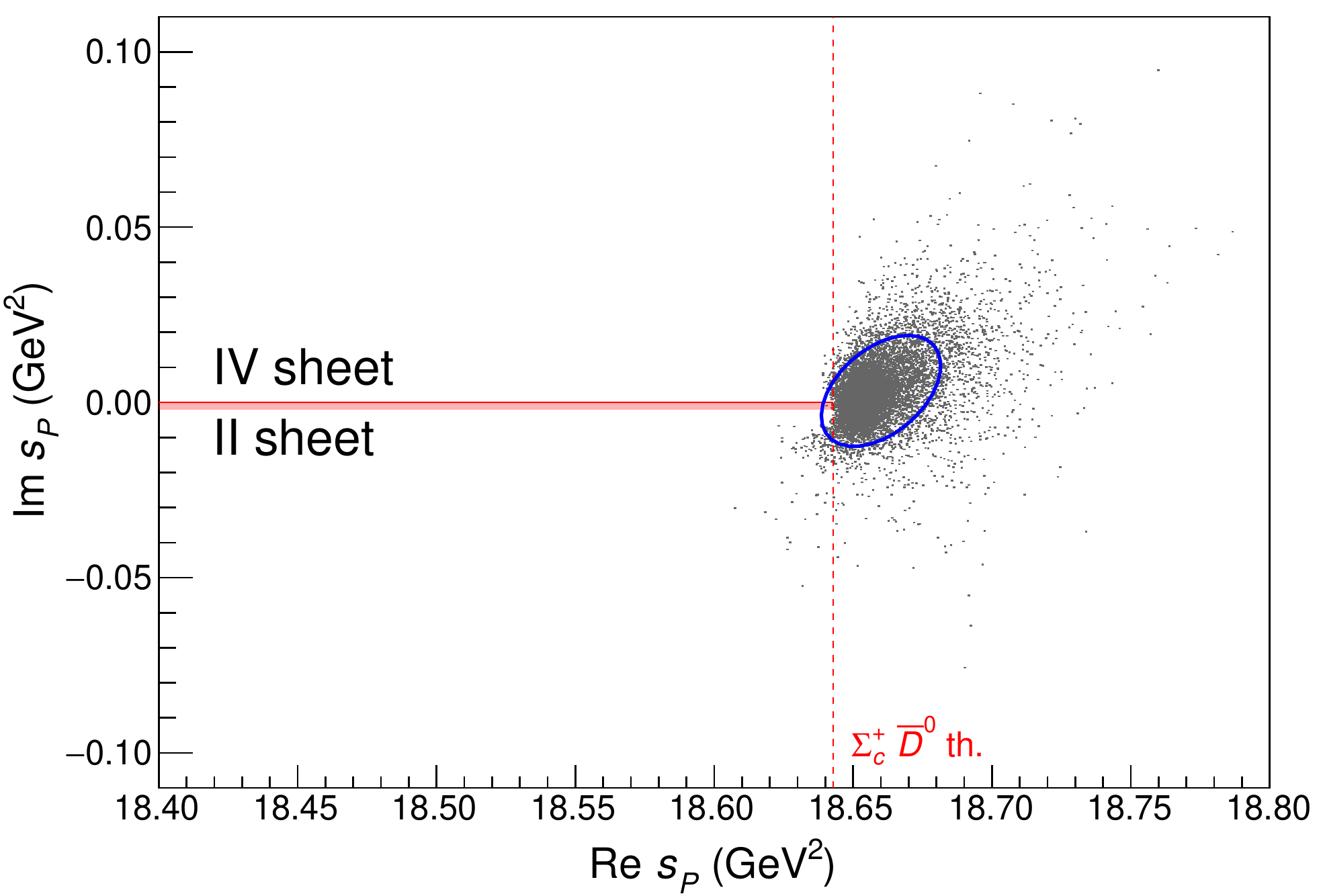}
\includegraphics[width=.49\textwidth]{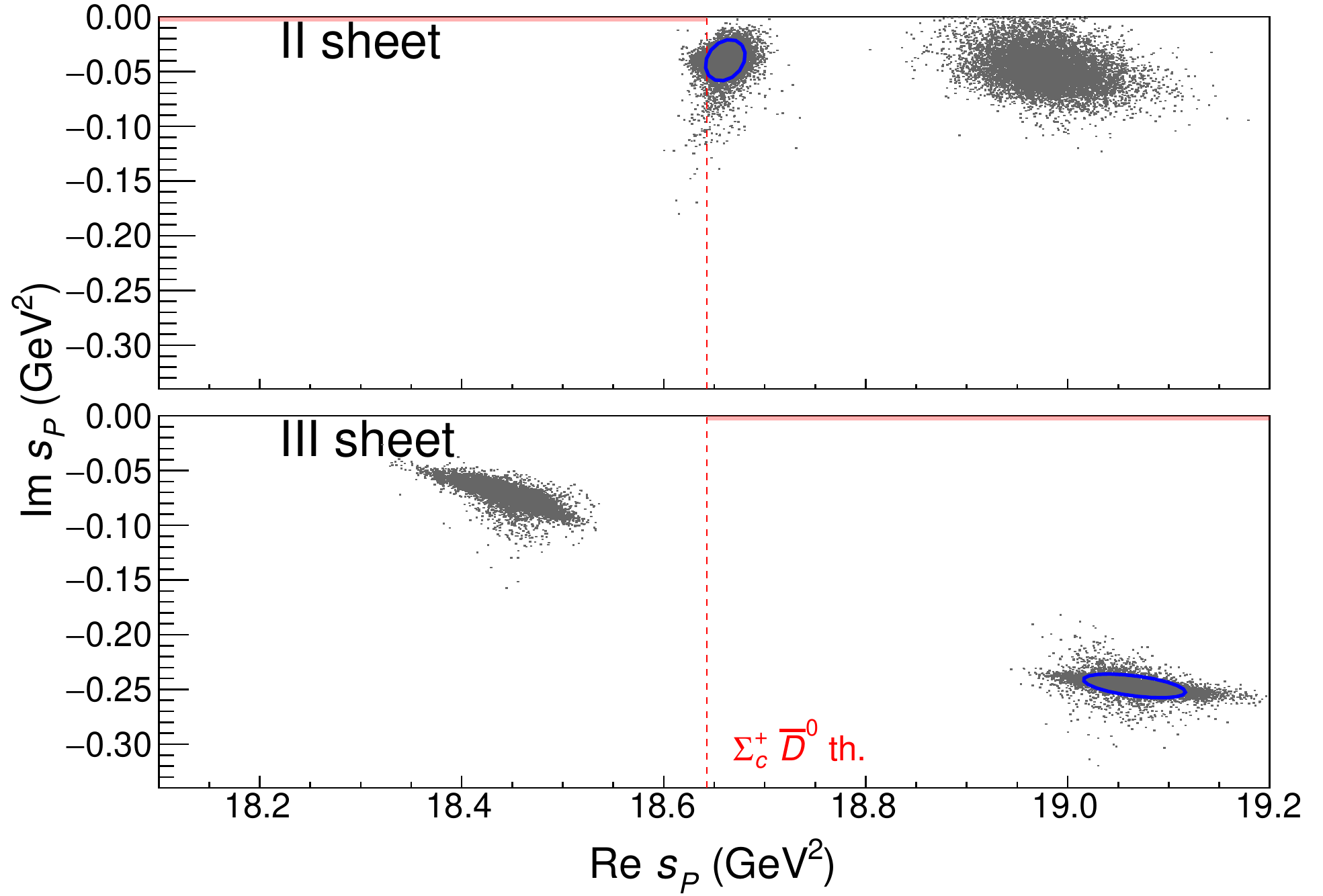}
\caption{
Poles obtained from the $10^4$ bootstrap fits
for cases~A (left) and~B (right). The physical region is highlighted with a pink band.
For case~A the poles lie on
the II and IV Riemann sheets
(which are continuously connected above the \SigmaD threshold).
For each bootstrap fit only one pole appears in this region 
and the blue ellipse accounts the 68\% of the cluster concentrating above threshold.
The right-hand plots show the poles for case~B. 
For each bootstrap fit we obtain a pole on the
II sheet and its partners on the III sheet.
The higher mass pole on the II sheet and its partner 
on the III sheet are above the fitted energy range and try
to capture the bumplike structure that appears at 
$4370\mev$. The lower mass pole on the II sheet and its partner
on the III sheet are responsible for the \Pc signal. 
The blue ellipses  account for 68\% of the two clusters.
} \label{fig:poles}
\end{figure*} 
Because of the square roots in $k_1$ and $k_2$, the amplitude has branch cuts opening at the two thresholds.
Through analytic continuation to complex values of $s$, one accesses four different Riemann sheets (see also Fig.~2 of Ref.~\cite{Pilloni:2016obd}). 
The physical region between the two thresholds is connected to the
  lower half of the II sheet. Similarly, the physical region above the \SigmaD threshold is connected to the 
  lower half of the III sheet. 
  Poles in these sheets will appear as peaks with Breit-Wigner-like line shape in data, if they lie
  below 
  the respective physical regions, \ie between the two thresholds for the II sheet, and above the heavier one for the III sheet.   
  From the II sheet, if one continuously moves to the upper 
   half plane above the higher threshold, one enters the upper half of the IV sheet. Since the latter is hidden from the physical region, a pole here will manifest in data as a cusp at the \SigmaD threshold. 

\mytitle{Results and discussion}
In order to determine the sensitivity of data to various scenarios, we consider two cases.
In case~A, we set $c_{ij}=0$, which corresponds to the scattering length approximation. This choice is substantially equivalent to the universal amplitude used in Ref.~\cite{Braaten:2007dw} to describe the $X(3872)$. It is known that the amplitude $T_{11}(s)$ 
 can have a pole on either the II or IV sheet, but not on the III sheet~\cite{Frazer:1964zz}. 
  This pole is entirely due to the opening 
    of the heavier channel, and therefore it
    is a measure of the strength of the \SigmaD interaction. 
Further interpretations can be drawn by changing the amplitude parameters and studying
  the movement of singularities, which in turn gives insight about their nature~\cite{Hanhart:2014ssa,Pelaez:2015qba}. For example, in potential scattering resonance poles appear for both attractive and repulsive potentials, but only the former move onto the real energy axis to  become bound states, when the interaction is strong enough.
In our case,   as the coupling between the two channels is turned off,  the pole could either move to the real axis of the physical sheet below the heavier threshold, thus representing a bound molecule, 
 or move onto the real axis of the unphysical sheet, corresponding to an unbound, virtual state.
  In case~B, we let the diagonal effective ranges $c_{ii}$ float. 
  The off-diagonal $c_{12}$ does not add other singularities, is not needed to describe data 
  and we set it
   to zero. In this case, poles related to the threshold as the ones just discussed are possible but not guaranteed; however, other poles can appear on the II and III sheet.\footnote{It is easy to check that case~A with $c_{ii} = 0$ has exactly 2 pairs of conjugate poles in the various sheets, while the general case~B has exactly 4. Only the closest to the physical region are relevant.} 
   The latter can be interpreted as originating from genuine pentaquark particles, with bare masses $\sqrt{m_{ii}/c_{ii}}$, that move into the complex plane and acquire a width when coupled to the open channels. The other clear distinction between these and the threshold-related poles discussed above is that the latter move far less in the complex plane when the channel couplings are varied.
  
We fit the data using~\textsc{MINUIT}~\cite{minuit}
and take into account the experimental resolution reported in~\cite{Aaij:2019vzc}.
The initialization of the parameters is chosen by randomly generating $O(10^5)$ different sets of values. The amplitude in Eq.~\eqref{1} is not protected against unphysical poles in the  I~sheet. Fits with such poles are discarded.
The best solutions for the two cases have comparable $\chi^2/\text{DOF}\simeq 0.8$. 
Figure~\ref{fig:fits} shows both fits to the data.
The preference of case~B over case~A is only at the $1.8\sigma$ level calculated with the Wilks theorem~\cite{Wilks:1938dza},
and we consider both cases as equally acceptable. 
In both cases, we find a pole $2\mev$ above the \SigmaD threshold, on the IV sheet for case~A and the II sheet for case~B. 
For case~B, additional poles appear 
  farther away from the \SigmaD threshold, on the II and III sheet. These do not affect the \Pc signal.

\begin{figure*}
\centering
\includegraphics[width=.49\textwidth]{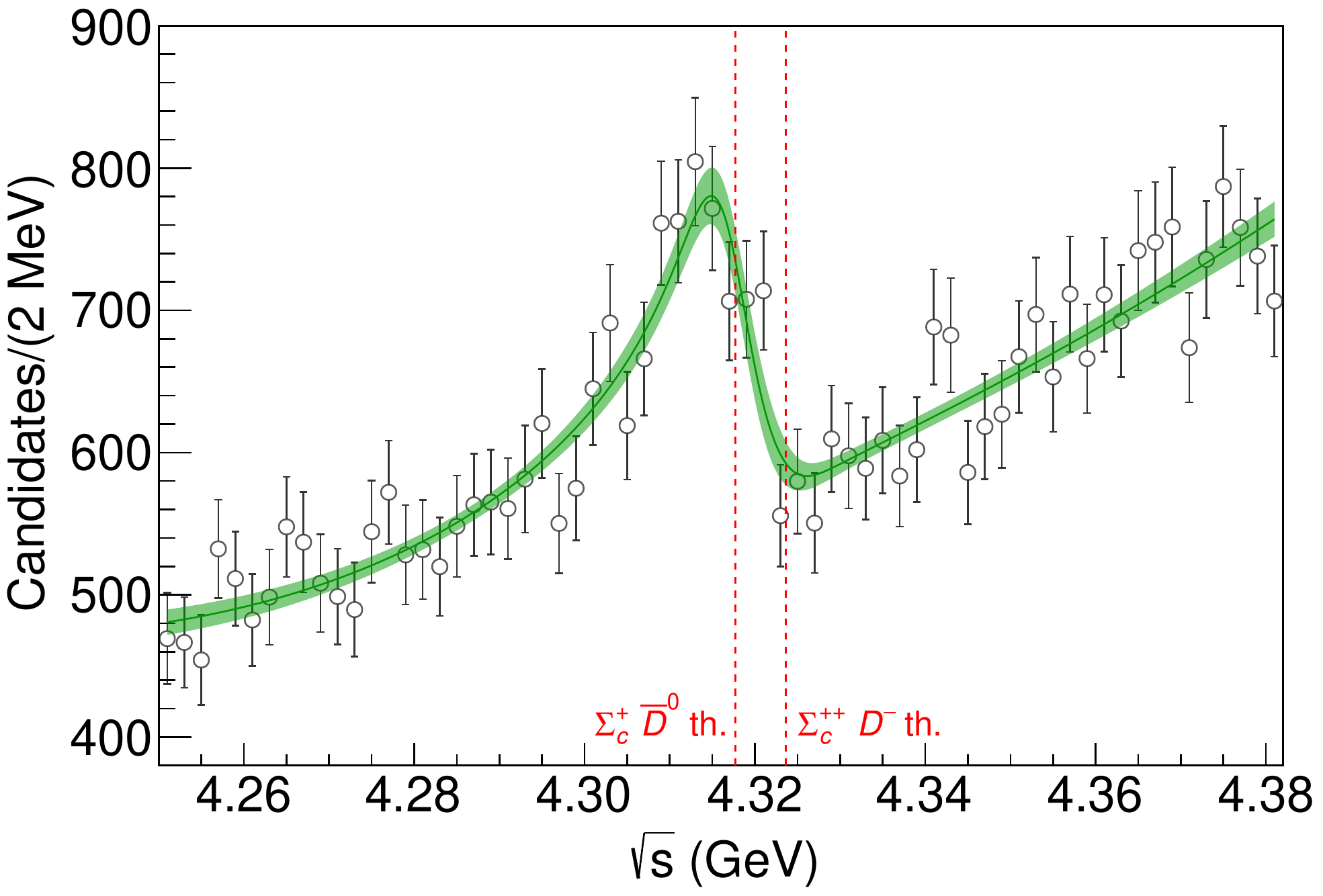}
\includegraphics[width=.49\textwidth]{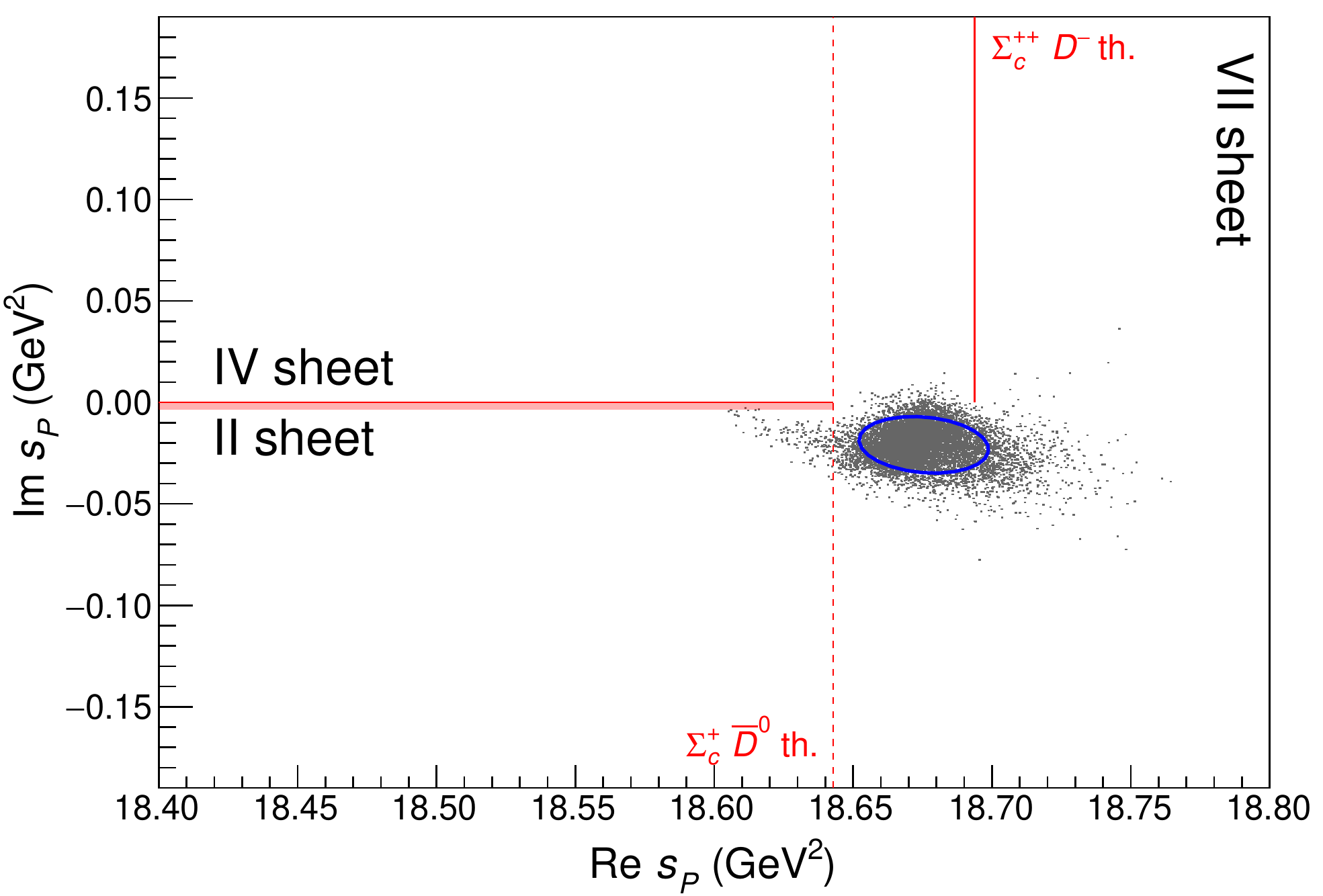}
\caption{
Fit to the $\cos\theta_{P_c}$-weighted \jpsip mass distribution (left) and pole positions (right) for the three-channel case.
Notation is the same as in Figs.~\ref{fig:fits} and~\ref{fig:poles}.
The poles are obtained from the $10^4$ bootstrap fits
and lie on the II and IV Riemann sheets
(which are continuously connected above the \SigmaD threshold)
and on the VII sheet (which is continuously connected to the II sheet
above the $\Sigma_c^{++} D^-$ threshold).
For each bootstrap fit only one pole appears in this region. The blue ellipse accounts for 68\% of the cluster concentrating above
the \SigmaD threshold.
} 
\label{fig:3chan}
\end{figure*} 
To estimate the sensitivity of the pole positions 
to the uncertainties in the data, we use the bootstrap technique~\cite{recipes,EfroTibs93}; 
\ie we generate $10^4$ pseudodata sets and fit each one of them. 
The statistical fluctuations in data reflect into the the uncertainty band plotted in Fig.~\ref{fig:fits}. Moreover, for each of these fits, we determine the pole positions, as shown in Fig.~\ref{fig:poles}. 

In case~A, it is possible to identify a cluster of virtual state poles across the II and IV sheet above the \SigmaD threshold (see also the discussion in Ref.~\cite{Rodas:2018owy}). 
If we use the customary definition of mass and width,
$M_P = \re \sqrt{s_p}$, $\Gamma_P = -2\im \sqrt{s_p}$  the main cluster has $M_P = 4319.7 \pm 1.6\mev$, $\Gamma_P = -0.8 \pm 2.4\mev$, where positive or negative values of the width correspond to II or IV sheet poles, respectively. To establish the nature of this singularity,  we track down the movement of the poles as 
the coupling between the two channels is reduced. By taking 
$m_{12} \to 0$, we can see how the cluster moves over to the upper side of the IV sheet and ends up 
on the real axis below the \SigmaD threshold~\cite{pcjpaclink}. The fraction of poles that reach the real axis from the lower side of the II sheet is $0.7\%$ only, and thus not significant.
This result reinforces the interpretation of the pole as an unbound virtual state, 
meaning that the binding between the $\Sigma_c^+$ and the $\bar D^0$ is insufficient to form a molecule. 

In case~B, the poles on the II sheet accumulate in two clusters. The one closer to threshold  has
$M_P = 4319.8 \pm 1.5\mev$ and $\Gamma_P = 9.2 \pm 2.9\mev$, and it is the one responsible for the \Pc signal. 
As we did for case~A, we study the motion of the poles
as the channels decouple. 
The lighter cluster migrates onto the IV sheet, where it hits the zero of $T_{11}(s)$, and annihilates when $m_{12} = 0$.
Since the pole does not survive the decoupling, it is entirely due to the interaction between the two channels, and its motion to the farthest unphysical sheet also suggests a virtual state nature for the \Pc as in case~A.

The other poles, which 
are located farther away from the \SigmaD threshold, can also be interpreted. 
There are two clusters of poles 
 on the III sheet, one far above the \SigmaD threshold, the other below (see Fig.\ref{fig:poles}). 
 The former could correspond to a resonance with standard Breit-Wigner line shape as it  appears to originate  
  from the  broad bump in the mass spectrum centered at $\sqrt{s} \sim 4.37\gev$.\footnote{ This is seen in 
   both weighted and unweighted datasets. 
In Ref.~\cite{Xiao:2019aya} a state is also found and identified with this bump. 
We also note that the $\Sigma_c(2520)\bar{D}^0$ threshold is close at $4383.24\mev$.} 
Although the presence of such a state is not significant, as is clear from the fact that case~A fits the data equally well, and no conclusion can be drawn before the complete amplitude analysis, it is an interesting speculation that such an enhancement might be related to the broad $P_c(4380)^+$ observed by the previous 
LHCb analysis~\cite{Aaij:2015tga}.

As $m_{12} \to 0$, the two III sheet clusters move close to each other, and the 
 heavier one disappears, multiplied 
  by the amplitude zeros when $m_{12} = 0$. In this uncoupled limit, only one pole per channel is left. 
 Furthermore, as  the channels close, 
which is achieved by replacing $ik_i \to \lambda i k_i$ and letting $\lambda \to 0$, the poles move onto the real axis. It is worth noting that the fit chooses almost identical values for the ratios 
$m_{11}/c_{11}\simeq m_{22}/c_{22}$. 
This ratio 
  determines the independent positions of the bare poles on the real axis  in the two uncoupled channels and being equal, suggests 
   existence of a single compact pentaquark.  

  We also performed a study of the 
   three-channel case, including the $\Sigma_c^{++}D^-$ threshold. 
   To simplify the approach we work in the scattering length approximation (as in case~A)
   and in the isospin limit for the fitting parameters. 
The result of the fit and the pole positions are shown in Fig.~\ref{fig:3chan} and details can be found in the Supplemental Material~\cite{pcjpaclink}.
We find a single pole close to the \SigmaD threshold on the II sheet. 
   No other pole appears close to the physical axis above the $\Sigma_c^{++}D^-$ threshold. 
     When the couplings between the channels are reduced, the
      pole quickly moves far to the left, and cannot be interpreted as a physical state. 
   We therefore conclude that the \Pc signal could be a result of a complicated interplay of thresholds and  feeble $\Sigma_c \bar D$ interactions. 
   
   We perform further systematic analyses by considering Flatt\'e and $K$-matrix parametrizations.  
Using a single $K$-matrix pole with an off-diagonal constant background leads to a pole on the II sheet in the same position as case~A. On the other hand, the Flatt\'e parametrization 
does not provide a good description of the \Pc peak, and does not generate stable poles in the region of interest.
 
As a cross-check we fit all the above approaches to the unweighted \jpsip spectrum in the same region, 
both with and without the $m_{Kp} > 1.9\gev$ cut. 
Results are consistent.

 \mytitle{Conclusions} 
  In summary, we have studied the \Pc reported by LHCb in the \jpsip spectrum. We considered a reaction amplitude which satisfies the general principles of $S$-matrix theory, with a minimum bias from the underlying theory. The analytic properties of the amplitudes can be related to the microscopic origin of the signal. We fitted the LHCb mass spectrum in the $4312\mev$ mass region including the experimental resolution. 
  The statistical uncertainties in the data were propagated to the extracted poles using the
  bootstrap technique. 
 We do not find support for a bound molecule. Based on 
 a systematic analysis of the reaction amplitudes, we conclude instead  
  that the interpretation of the \Pc peak 
   as a virtual (unbound) state is more likely.

\mytitle{Acknowledgments} 
We thank Tomasz Skwarnicki for discussions  and useful comments on the manuscript.
This work was supported by
the U.S.~Department of Energy Grants
No.~DE-AC05-06OR23177 
and No.~DE-FG02-87ER40365, 
U.S.~National Science Foundation Grant No.~PHY-1415459, 
PAPIIT-DGAPA (UNAM, Mexico) Grant No.~IA101819, 
and CONACYT (Mexico) Grants No.~734789, No.~251817,
and~No.~A1-S-21389. 
V.M. acknowledges support from Comunidad Aut\'onoma de Madrid through 
Programa de Atracci\'on de Talento Investigador 2018.


\bibliographystyle{apsrev4-1-jpac}
\bibliography{quattro.bib}

\clearpage
\onecolumngrid
\section{Supplemental material}
\begin{itemize}
\item The animations of the pole motion are available in GIF format on \href{http://www.indiana.edu/~jpac/pc4312.php}{http://www.indiana.edu/$\sim$jpac/pc4312.php}.
\item We write here the explicit formula for the amplitudes described in the text. We use isospin to relate the \SigmaD and the $\Sigma_c^{++} D^-$ channels: 
\begin{align*}
\frac{dN}{d\sqrt{s}} & = \rho(s) \left[ \abs{F(s)}^2   
 + B(s) \right],\\
F(s) &= P_1(s) T_{11} (s),
\end{align*}%
\begin{equation*}
T(s) = \begin{pmatrix}
m_{11} - c_{11} s - i \sqrt{s - s_1} & m_{12} & \xi\, m_{12}\\ 
m_{12} & m_{22} - c_{22} s - i \sqrt{s - s_2} & \xi\, m_{23} & \\
\xi \, m_{12} & \xi \, m_{23} & 1 + \xi \left(-1 + m_{22} - c_{22} s - i \sqrt{s - s_3}\right)\end{pmatrix}^{-1},
\end{equation*}
with $s_1 = \left(m_{\psi} + m_{p}\right)^2$, $s_2 = \left(m_{\Sigma_c^{+}} + m_{\bar D^0}\right)^2$, and $s_3 = \left(m_{\Sigma_c^{++}} + m_{D^-}\right)^2$.
If $\xi=0$ the amplitude $T(s)$ reduces to Eq.~\eqref{eq:caseAB}.
\end{itemize}
\begin{table}[h]
\caption{Summary of fit results for the three cases described in the text. 
Appropriate powers of \gev units are understood.
   The function $P_1(s)$ is parameterized as $p_0 + p_1 s$
   and $B(s)$ as $b_0 + b_1 s$.
In each case the  $\chi^2/\text{dof}$ corresponds to the best fit obtained.
The first column of parameters is obtained from the best fit, and should be used to reproduce the plots. The second column reports 
   the mean value and uncertainty of the parameters from bootstrap.
The phase space factor for the decay $\Lambda_b^0 \to \jpsip\, K^-$ appearing 
 in Eq.~\eqref{1} is given by $\rho(s) =m_{\Lambda_b} p\, q$ 
 with 
 $p = \lambda^{1/2}(s,m^2_{\Lambda_b},m^2_K)/2m_{\Lambda_b}$, $q = \lambda^{1/2}(s,m^2_p,m^2_\psi)/2\sqrt{s}$,
 and $\lambda(x,y,z) = x^2 + y^2 + z^2 - 2xy - 2xz - 2yz$ is the K\"all\'en function. The size and shape of the incoherent  background $B(s)$ are comparable for all the cases, despite large variations in $b_0$ and $b_1$. The $P(s)$ polynomials are also similar up to a scale factor. This gives further support that changing amplitudes as described in the text probes the dynamics of final state interactions.
}
\begin{ruledtabular}
\label{table} 
\begin{tabular}{c|cc|cc|cc}
& \multicolumn{2}{c|}{Case~A} 
& \multicolumn{2}{c|}{Case~B}  
&\multicolumn{2}{c}{3-channel} \\
\hline
$\chi^2/\text{dof}\phantom{0}$&\multicolumn{2}{c|}{$48.1/(66-7)=0.82$} & \multicolumn{2}{c|}{$43.0/(66-9)=0.75$} & \multicolumn{2}{c}{$45.5/(66-8)=0.78$}\\
\hline
& best fit & bootstrap & best fit& bootstrap& best fit& bootstrap\\
\hline
$b_0$ &$402.95$&$446\pm73 $&$0.74$& $6.1\pm 6.0$  &$121.56 $&$123.1\pm1.4$ \\
$b_1$ &$-15.00$&$-17.4\pm4.1 $ &$7.22$& $6.93\pm0.36$  &$0.63$&$0.52\pm0.14$ \\
$p_0$ &$423.16$& $437\pm16 $ &$85.06$&$92.6\pm 8.8$ &$422.72$&$422.52\pm0.38$ \\
$p_1$ &$-23.53$&$-24.28\pm0.81$&$-5.30$& $-5.70\pm0.47$&$-23.41$& $-23.409\pm0.040$ \\
$m_{11}$ &$2.60$&$2.65\pm0.28$&$151.29$&$151.35\pm0.23$&$2.83 $& $2.82\pm0.19$  \\
$m_{22}$ &$0.22$& $0.223\pm0.078$&$38.81$&$39.12\pm0.28$ &$-4.27$&$-4.259\pm0.042$  \\
$m_{12}$ &$0.85$&$0.86\pm0.11$ &$1.03 $&$1.035\pm0.062$ &$0.64$&$0.646\pm0.057$ \\
$m_{23}$ &$0$&$0$ &$0 $&$0$ &$4.38$& $4.385\pm0.022$\\
$c_{11}$  &$0$&$0$ &$8.00$&$8.007\pm0.015$ &$0$& $0$\\
$c_{22}$  &$0$&$0$ &$2.06$&$2.081\pm0.016$ &$0$&$0$\\
$c_{12}$  &$0$&$0$ &$0$&$0$ &$0$&$0$\\
$\xi$  &$0$&$0$ &$0$&$0$ &$1$&$1$\\

\end{tabular}
\end{ruledtabular}
\end{table}

\begin{figure*}[h]
\centering
\includegraphics[width=.49\textwidth]{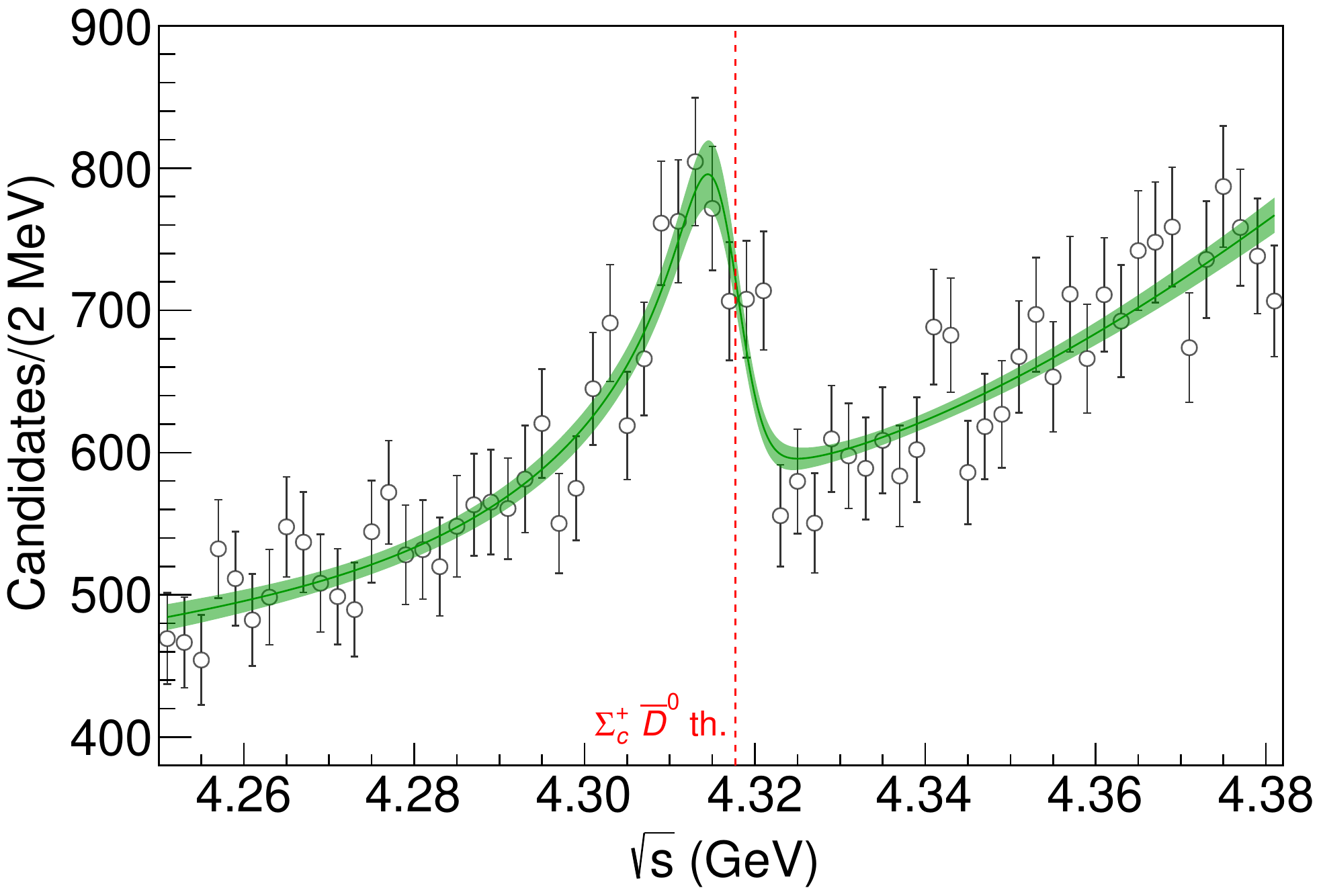}
\includegraphics[width=.49\textwidth]{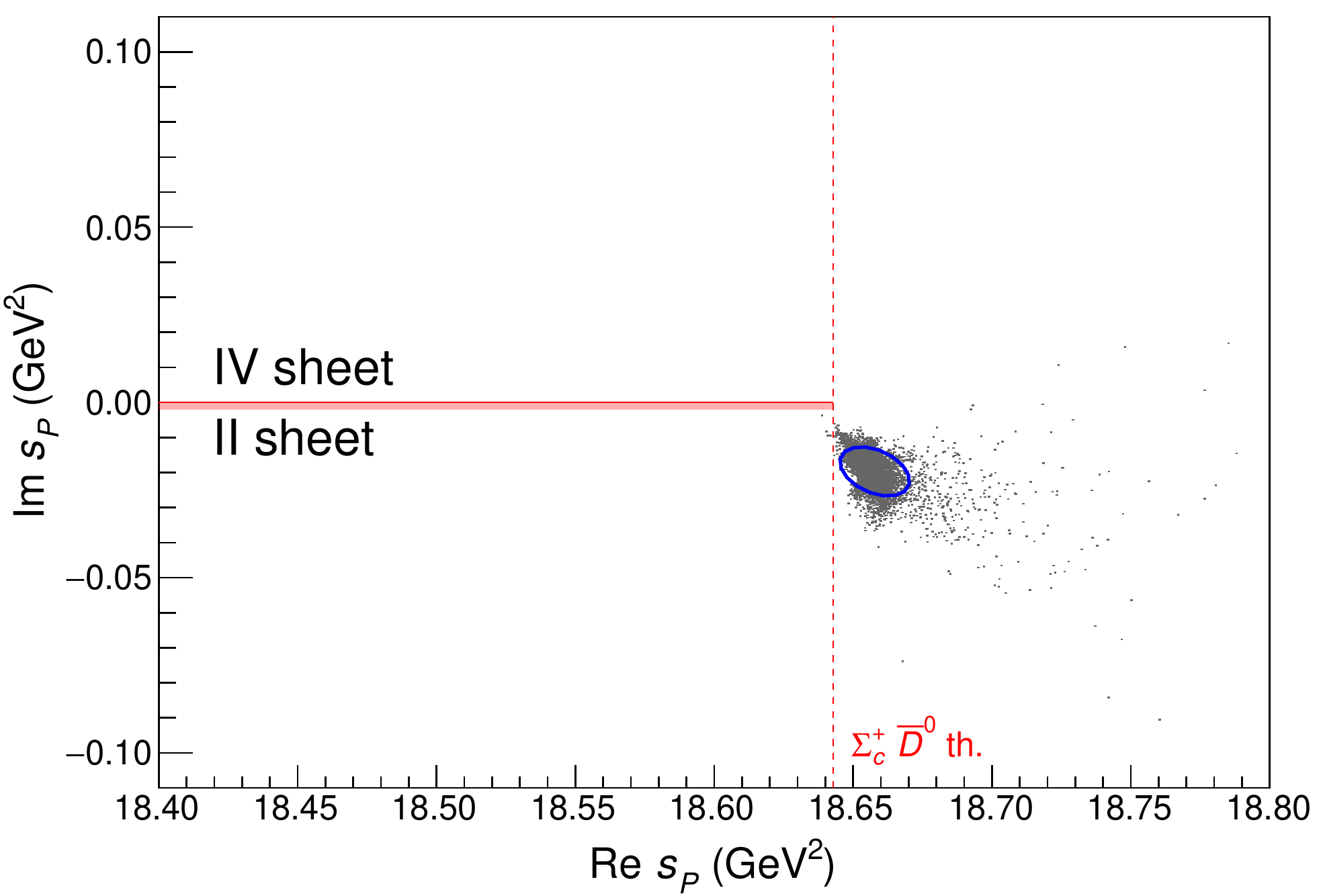}
\caption{
Fit to the $\cos\theta_{P_c}$-weighted \jpsip mass distribution (left) and pole positions (right) for the $K$-matrix case.
Notation is the same as in Figs.~\ref{fig:fits} and~\ref{fig:poles}.
The poles are obtained from the $10^4$ bootstrap fits
and lie on the II and IV Riemann sheets
(which are continuously connected above the \SigmaD threshold).
} \label{fig:kmat}
\end{figure*} 

\begin{figure*}[h]
\centering
\includegraphics[width=.49\textwidth]{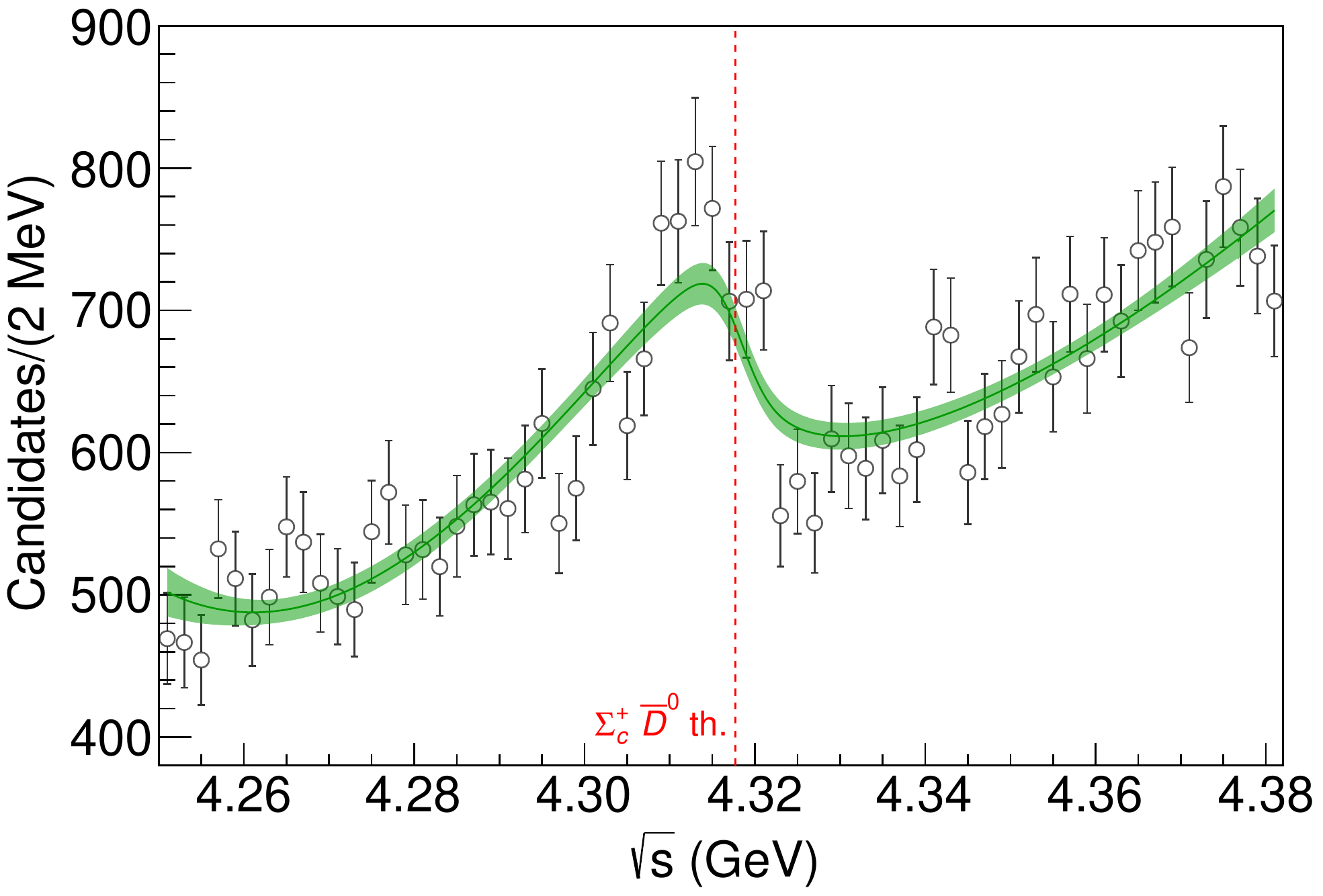}
\caption{Fit to the $\cos\theta_{P_c}$-weighted \jpsip mass distribution for the Flatt\'e case.
Notation is as in Fig.~\ref{fig:fits}.
This parameterization does not generate a pole in the region of interest.
} \label{fig:flatte}
\end{figure*} 

\end{document}